\documentclass{article}

\usepackage{graphicx}
\usepackage{amsmath}
\usepackage{relsize}

\usepackage{fancyheadings}
\newcommand{\panda}{${\sf \overline{P}ANDA}$\hspace*{1ex}}
\newcommand{\BaBar}{\mbox{\slshape B\kern -0.1em{\smaller A}\kern-0.1em B\kern-0.1em{\smaller A\kern -0.2em R }}}

\renewcommand{\Gamma}{\varGamma}

\begin{document}

\parindent0em

\title{XYZ States - Results from Experiments}

\author{{\slshape S\"oren Lange}\\[1ex]
Justus-Liebig-Universit\"at Giessen, II.~Physikalisches Institut\\
Heinrich-Buff-Ring 16, 35392 Giessen, Germany}



\maketitle

\noindent
\begin{center}
{\sf Lecture given at the}\\ 
{\sf Helmholtz International Summer School Physics of Heavy Quarks and Hadrons}\\
{\sf Dubna, Russia, 07/15-28/2013.}
\end{center}

\section{Introduction}

The static quark anti-quark potential in strong interaction 
is often expressed using the ansatz 

\begin{eqnarray}
V(r) & = & - \frac{4}{3} \frac{\alpha_s}{r} + k r \nonumber\\
& & + \frac{32 \pi \alpha_s}{9 m_c^2}
\delta ( r ) \vec{S_c} \vec{S_{\overline{c}}} \nonumber\\
& & + \frac{1}{m_c^2}
( \frac{2 \alpha_s}{r^3} - \frac{k}{2 r} ) \vec{L} \vec{S}\nonumber\\
& & + \frac{1}{m_c^2}
\frac{4 \alpha_s}{r^3}
( \frac{3 \vec{S_c} \vec{r} \cdot \vec{S_{\overline{c}}} \vec{r} }
{r^2}
- \vec{S_c} \vec{S_{\overline{c}}} ) \quad .
\label{ecornell}
\end{eqnarray}

For historic reasons, this potential is refered to as 
a Cornell-type potential
\cite{potential_1978}
\cite{potential_1985}
\cite{potential_2005}.
The first term is a Coulomb-like term describing one-gluon
exchange, which is very similar to the Coulomb term in QED 
potentials for e.g.\ positronium or the hydrogen atom, 
except that here the coupling constant is given by $\alpha_S$
instead of $\alpha_{em}$. 
The second term is a linear term which phenomenologically 
describes QCD confinement, and which is completely absent 
in QED. The linear shape is e.g.\ supported by Lattice 
QCD calculations, and the parameter $k$ is the string 
constant of QCD string between the quark and the anti-quark. 
The other terms represent spin-orbit, spin-spin and tensor
potentials, leading to mass splittings in the spectrum.

Heavy quark combinations such as the charm anti-charm (called 
charmonium) and the beauty anti-beauty (called bottomonium)
are in particular interesting, as they can be treated 
{\it (a)} as non-relativistic systems and 
{\it (b)} perturbatively due to $m_Q$$>>$$\Lambda_{QCD}$, 
where $\Lambda_{QCD}$$\simeq$200~MeV is the QCD scale.

Charmonium- and bottomonium spectroscopy has been a flourishing 
field recently, as many new states have been observed.
Masses of expected states (such as the $h_b$, $h_b'$, $\eta_b$, $\eta_b'$, 
described below) have been measured accurately and enable 
precision tests of Eq.~\ref{ecornell}
to a level of $\Delta$$m$/$m$$\leq$10$^{-4}$.

On the other hand, several non-expected states were found, 
which do not fit into the Cornell-type potential model prediction. 
While for many priorly observed charmonium and bottomonium states
the difference between predicted and measured mass is impressively small
in the order of $\Delta$$m$$\simeq$2-3~MeV, for some of the new 
states the closest predicted state is off by $\Delta$$m$$\geq$50~MeV
or more. Such states are often refered to as XYZ states. 
The Z states (as will be described below) are in particular interesting, 
as they are charged states, and thus can not represent charmonium or
bottomonium at all. 

Many of the $XYZ$ states were observed at the Belle \cite{belle_nim}
and BaBar \cite{babar_nim} experiments in $e^+$$e^-$ collisions 
at beam energies 10.5-11.0~GeV (i.e.\ in the $\Upsilon$(nS) region). 
In this draft, at first charmonium-like states will be discussed, 
which are e.g.\ produced in $B$ meson decays. 
Belle and \BaBar are often called $B$ meson factories,
as the number of produced $B$ mesons per time unit is very high. 
Often the size of a data sample is given as integrated lumnosity.
With a typical instantenous luminosity of 1$\times$10$^{34}$~s$^{−1}$~cm$^{−2}$
and using 1~b(''barn'')=10$^{−24}$~cm$^2$, we get $\simeq$1$\times$10$^{-15}$~b$^{-1}$
or $\simeq$1~fb$^{-1}$ per 1 day. 
The center-of-mass energy of Belle and BaBar is $\sqrt{s}$=10.58~GeV,
corresponding to the mass of the $\Upsilon$(4S) resonance.
The cross section is $\sigma$($e^+$$e^-$$\rightarrow$$\Upsilon$(4S))$\simeq$1~nb,
and thus we get about 1$\times$10$^6$ produced $B$ meson pairs per day. 

Further below in this draft, examples for bottomonium-like states will be given, 
which are e.g.\ produced in radiative decays of 
$\Upsilon($n$S)$ resonances. 
As an example of applications of the measurements, 
a few precision tests of the Cornell-type potential (Eq.~\ref{ecornell})
will be discussed. 
At the end, an outlook to a future experiment will be given, which will be able
to measure the width of a state in the sub-MeV regime. 

\section{Charmonium(-like) states}

\subsection{The X(3872) state}

The X(3872) state has been discovered in $B$ meson decay
in the decay X(3872) $\rightarrow$$J$/$\psi$$\pi^+$$\pi^-$
by Belle \cite{x3872belle} and confirmed by other experiments 
\cite{x3872babar} 
\cite{x3872cdf} 
\cite{x3872d0}
\cite{x3872lhcb}
\cite{x3872cms}.
Among the XYZ states, 
the X(3872) is the only one observed in several decay channels: 
X(3872)$\rightarrow$$J$/$\psi$$\pi^+$$\pi^-$,
X(3872)$\rightarrow$$J$/$\psi$$\gamma$, 
X(3872)$\rightarrow$$J$/$\psi$$\pi^+$$\pi^-$$\pi^0$, 
X(3872)$\rightarrow$$D^0$$\overline{D}^0$$\pi^0$, and 
X(3872)$\rightarrow$$D^0$$\overline{D}^0$$\gamma$. 

\begin{figure}[tbh]
\includegraphics[width=\textwidth,height=3cm]{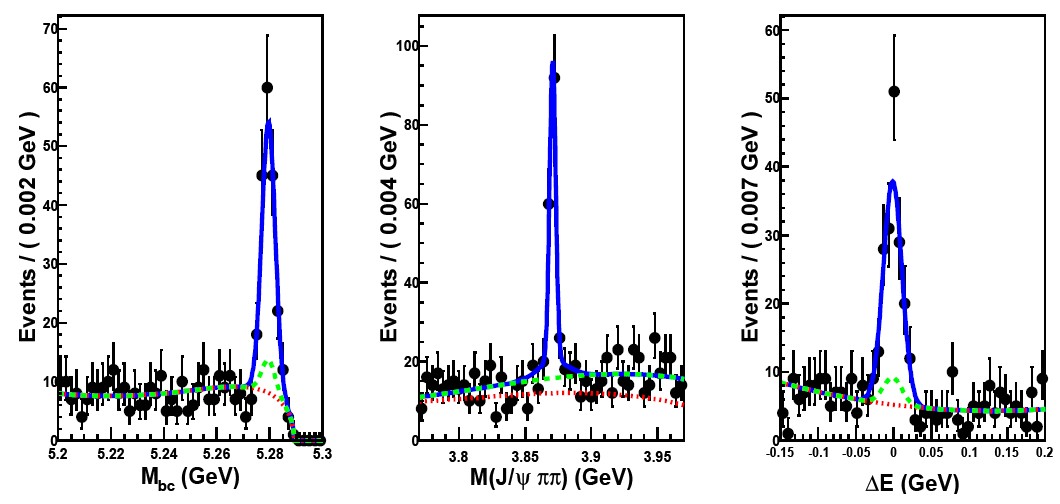}
\caption{Beam constrained mass $M_{\rm bc}$=$\sqrt{(E_{beam}^{cms}/2)^2-(p_B^{cms})^2}$ (left)
invariant mass $m$($J$/$\psi$$\pi^+$$\pi^-$) 
and the energy difference $\Delta$$E$=$E_B^{cms}$$-$$E_{beam}^{cms}$ 
for the decay $B^+$$\rightarrow$$K^+$X(3872)($\rightarrow$$J$/$\psi$$\pi^+$$\pi^-$).
A 3-dimensional fit is performed. The blue line represents the fit result,
which is used to extract the mass and the width of the X(3872).
\label{fx3872}}
\end{figure}

\noindent
The mass of the X(3872) can be determined with high precision. 
A recent mass measurement of the X(3872) at Belle was 
based upon the complete Belle data set of 711~fb$^{-1}$
collected at the $\Upsilon$(4S) resonance. 
Fig.~\ref{fx3872}
shows the beam constrained mass $M_{\rm bc}$=$\sqrt{(E_{beam}^{cms}/2)^2-(p_B^{cms})^2}$ 
(left, with the energy in the center-of-mass system $E_{beam}^{cms}$ and the momentum
of the $B$ meson in the center-of-mass system $p_B^{cms}$), 
the invariant mass $m$($J$/$\psi$$\pi^+$$\pi^-$) (center)
and the energy difference $\Delta$$E$=$E_B^{cms}$$-$$E_{beam}^{cms}$ 
(right, with the energy of the $B$ meson in the center-of-mass system $E^{cms}_B$).
Data and fit (as a result of a 3-dimensional fit to the observables 
shown) for the decay 
$B^+$$\rightarrow$$K^+$X(3872)($\rightarrow$$J$/$\psi$$\pi^+$$\pi^-$)
are shown (blue line: signal, dashed green line: background).
The fitted yield is 151$\pm$15 events. 
For details of the analysis procedure see \cite{x3872belle_width_2011}.
The fitted mass is 
listed in Tab.~\ref{tx3872} in comparison with mass
measurements from other experiments. 

The mass measurement reveals the surprising fact that the X(3872) 
is very close to the $D^{*0}$$\overline{D}^0$ threshold. 
Therefore it was discussed, if the X(3872) possibly represents 
an $S$-wave $D^{*0}$$\overline{D}^0$
molecular state \cite{tornqvist}.
In this case, the binding energy $E_b$ would be given 
by the mass difference $m$(X)$-$$m$($D^{*0}$)$-$$m$($D^0$). 
Including the new Belle result, the new world average mass of the X(3872)
is $m$=3871.68$\pm$0.17~MeV \cite{pdg}.
The present value for the sum of the masses is 
$m$($D^0$)+$m$($D^{*0}$)=3871.84$\pm$0.28~MeV \cite{pdg},
Thus, a binding energy of \mbox{$E_b$=$-$0.16$\pm$0.33~MeV} can be calculated,
which is enormously small. 
In addition, $E_b$ is inverse proportional to the squared
scattering length $a$ \cite{braaten_qwg07}:

\begin{equation}
E_b = \frac{\hbar^2}{2 \mu a^2}
\end{equation}

using the reduced mass $\mu$.
The radius can in first order be approximated by $<$$r$$>$=$a$/$\sqrt{2}$.
This would surprisingly mean a very large radius
$<$$r$$>$$\geq$10$^{+\infty}_{-5}$~fm of the mo\-le\-cu\-lar state. 

\begin{table}[tbh]
\begin{center}
\begin{tabular}{|l|l|l|}  
\hline
Experiment & Mass of X(3872) & \\
\hline
\hline
CDF2 & 3871.61$\pm$0.16$\pm$0.19~MeV & \cite{x3872cdf} \\
\hline
BaBar ($B^+$) & 3871.4$\pm$0.6$\pm$0.1~MeV & \cite{x3872babar} \\
\hline
BaBar ($B^0$) & 3868.7$\pm$1.5$\pm$0.4~MeV & \cite{x3872babar} \\
\hline
D0 & 3871.8$\pm$3.1$\pm$3.0~MeV & \cite{x3872d0} \\
\hline
Belle & 3871.84$\pm$0.27$\pm$0.19~MeV & \cite{x3872belle_width_2011} \\
\hline
LHCb & 3871.95$\pm$0.48$\pm$0.12~MeV & \cite{x3872lhcb} \\
\hline
\hline
New World Average & 3871.68$\pm$0.17~MeV & \cite{pdg} \\
\hline
\end{tabular}
\caption{Mass measurements of the X(3872).\label{tx3872}}
\end{center}
\end{table}

An important decay of the X(3872) is the radiative decay
X(3872)$\rightarrow$$J$/$\psi$$\gamma$.
The observation of this decay was reported by Belle
with a data set of 256~$fb^{-1}$,
a yield of 13.6$\pm$4.4 events
and a statistical significance of 4.0$\sigma$ \cite{x3872jpsigamma_belle}.
The combined branching ratio was measured to
$BR$($B^{\pm}$$\rightarrow$$X$$K^{\pm}$,
$X$$\rightarrow$$\gamma$$J$/$\psi$)=
(1.8$\pm$0.6$\pm$0.1)$\times$10$^{-6}$),
i.e.\ the branching fraction of X(3872)$\rightarrow$$J$/$\psi$$\gamma$ is
a factor $\simeq$6 smaller than the one for X(3872)$\rightarrow$$J$/$\psi$$\pi^+$$\pi^-$,
and thus this decay represents a rare decay. 
However, the decay is very important, as it 
represents a decay into two neutral particles,
which are identical to their anti-particles. 
Therefore observation of the decay implies, that the charge conjugation
of the X(3872) must be C=$+$1.
\BaBar was able to confirm the observation
with a data set of 260~$fb^{-1}$,
a yield of 19.4$\pm$5.7 events
and a statistical significance of 3.4$\sigma$ \cite{x3872jpsigamma_babar}.
Charmonium states with C=$+$1 are interesting objects. 
While decay widths (which can be measured by branching fractions in the experiment)
for C=$-$1 states scale with the squared modulus of the wave function
($\Gamma$$\sim$$|$$\Psi$($r$=0)$|^2$), decay widths of C=$+$1 states
scale with the squared modules of the {\it derivative} of the wave function
($\Gamma$$\sim$$|$$\partial$$\Psi$/$\partial$$r$($r$=0)$|^2$). 

An additional surprising property of the X(3872) is isospin violation.
It was found, that in the decay X(3872)$\rightarrow$$J$/$\psi$$\pi^+$$\pi^-$ 
the invariant mass peaks at the mass of the $\rho^0$ meson. 
The $\rho^0$ carries isospin $I$=0, but the initial state (if assumed 
to be a pure $c$$\overline{c}$ state) has $I$=0 (as it would not contain
any $u$ or $d$ valence quarks). 
There are only two additional isospin violating transitions known 
in the charmonium system \cite{pdg}, namely
$\psi'$$\rightarrow$$J$/$\psi$$\pi^0$ 
(${\cal B}$=1.3$\pm$0.1$\cdot$10$^{-3}$)
and $\psi'$$\rightarrow$$h_c$$\pi^0$ 
(${\cal B}$=8.4$\pm$1.6$\cdot$10$^{-4}$).
These branching fractions are very small. 
One of the mechanisms to induce isospin violation 
is the $u$/$d$ quark mass difference in strong interaction.
However, as the mass difference is small, the effect should be very small, 
consistent with the the measured branching fractions. 
Another possible mechanism to induce isospin violation is 
the $u$/$d$ quark charge difference in electromagnetic interactions (EM).
Isospin should only be conserved in strong interaction,
but not in EM interaction. Thus one of the possible
explanations might be, that the decay
X(3872)$\rightarrow$$J$/$\psi$$\rho$($\rightarrow$$\pi^+$$\pi^-$)
is proceeding via EM interaction, i.e.\ the $\rho$
might not be created be two gluons, but by a virtual photon.
However, then the decay should be suppressed by an additional
factor $\alpha_{em}$/$\alpha_S$$\simeq$10. 
The observation for the X(3872) is different:
the branching fraction of isospin violating 
transistion is (among the known decays) 
order of {\cal O}(10\%) and thus seems to be largely enhanced.

{\subsection{The Y(4260) family}

\begin{table}[htb]
\begin{center}
\begin{footnotesize}
\begin{tabular}{|l|l|l|l|l|l|l|}
\hline
 & 
BaBar \cite{y4260_babar_1} & 
CLEO-c \cite{y4260_cleo-c} & 
Belle \cite{y4260_belle_1} & 
Belle \cite{y4260_belle_2}  &
BaBar \cite{y4260_babar_2} &
BaBar \cite{y4260_babar_3} \\
\hline
${\cal L}$ &
211~fb$^{-1}$ &
13.3~fb$^{-1}$ &
553~fb$^{-1}$ &
548~fb$^{-1}$ &
454~fb$^{-1}$ &
454~fb$^{-1}$ \\
\hline
N & 
125$\pm$23 & 
14.1$^{+5.2}_{-4.2}$ & 
165$\pm$24 & 
324$\pm$21 &
344$\pm$39 &
$-$ \\
\hline
Significance & 
$\simeq$8$\sigma$ & 
$\simeq$4.9$\sigma$ & 
$\geq$7$\sigma$ & 
$\geq$15$\sigma$ &
$-$ &
$-$\\ 
\hline
$m$ / MeV & 
4259$\pm$8$^{+2}_{-6}$ & 
4283$^{+17}_{-16}$$\pm$4 & 
4295$\pm$10$^{+10}_{-3}$ & 
4247$\pm$12$^{+17}_{-32}$ &
4252$\pm$6$^{+2}_{-3}$ &
4244$\pm$5$\pm$4 \\
\hline
$\Gamma$ / MeV & 
88$\pm$23$^{+6}_{-4}$ & 
70$^{+40}_{-25}$ & 
133$\pm$26$^{+13}_{-6}$ & 
108$\pm$19$\pm$10 &
105$\pm$18$^{+4}_{-6}$ &
114$^{+16}_{-15}$$\pm$7 \\
\hline
\end{tabular}
\end{footnotesize}
\end{center}
\caption{Summary of the mass and width measurements of the Y(4260).\label{ty4260}}
\end{table}


\begin{figure}[hhh]
\centerline{\includegraphics[width=1.0\textwidth]{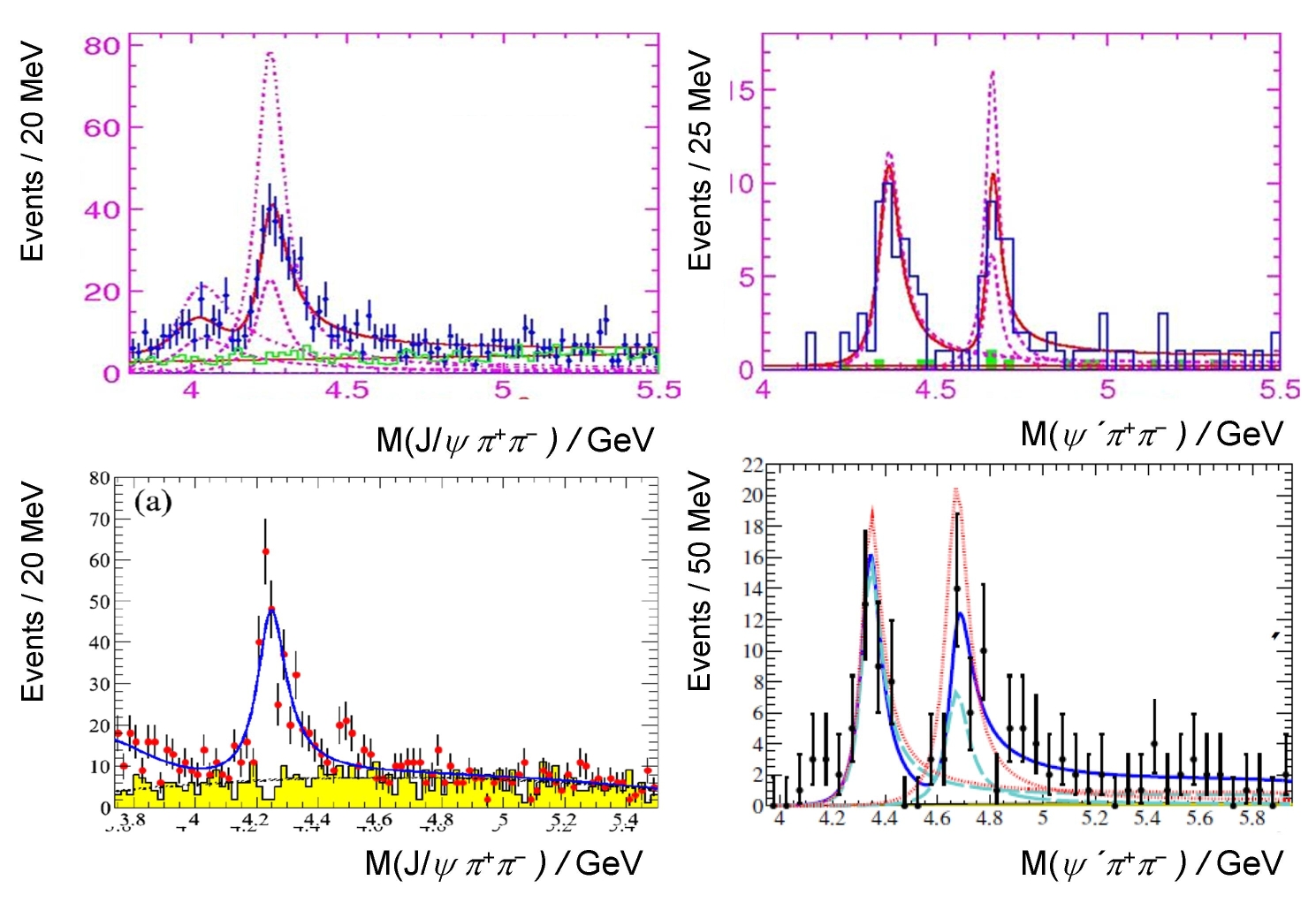}}
\caption{Observations of the Y states. 
Invariant mass $m$($J$/$\psi$$\pi^+$$\pi^-$) 
at Belle \cite{y4260_belle_2} (top left) 
and at BaBar \cite{y4260_babar_2} (bottom left).
Invariant mass $m$($\psi'$$\pi^+$$\pi^-$) 
at Belle \cite{y4350_belle} (top right) and 
at BaBar \cite{y4350_babar} (bottom right). Different curves indicate
different fits with or without interference.\label{fy4260}}
\end{figure}


\begin{figure}[hhh]
\centerline{\includegraphics[width=0.8\textwidth]{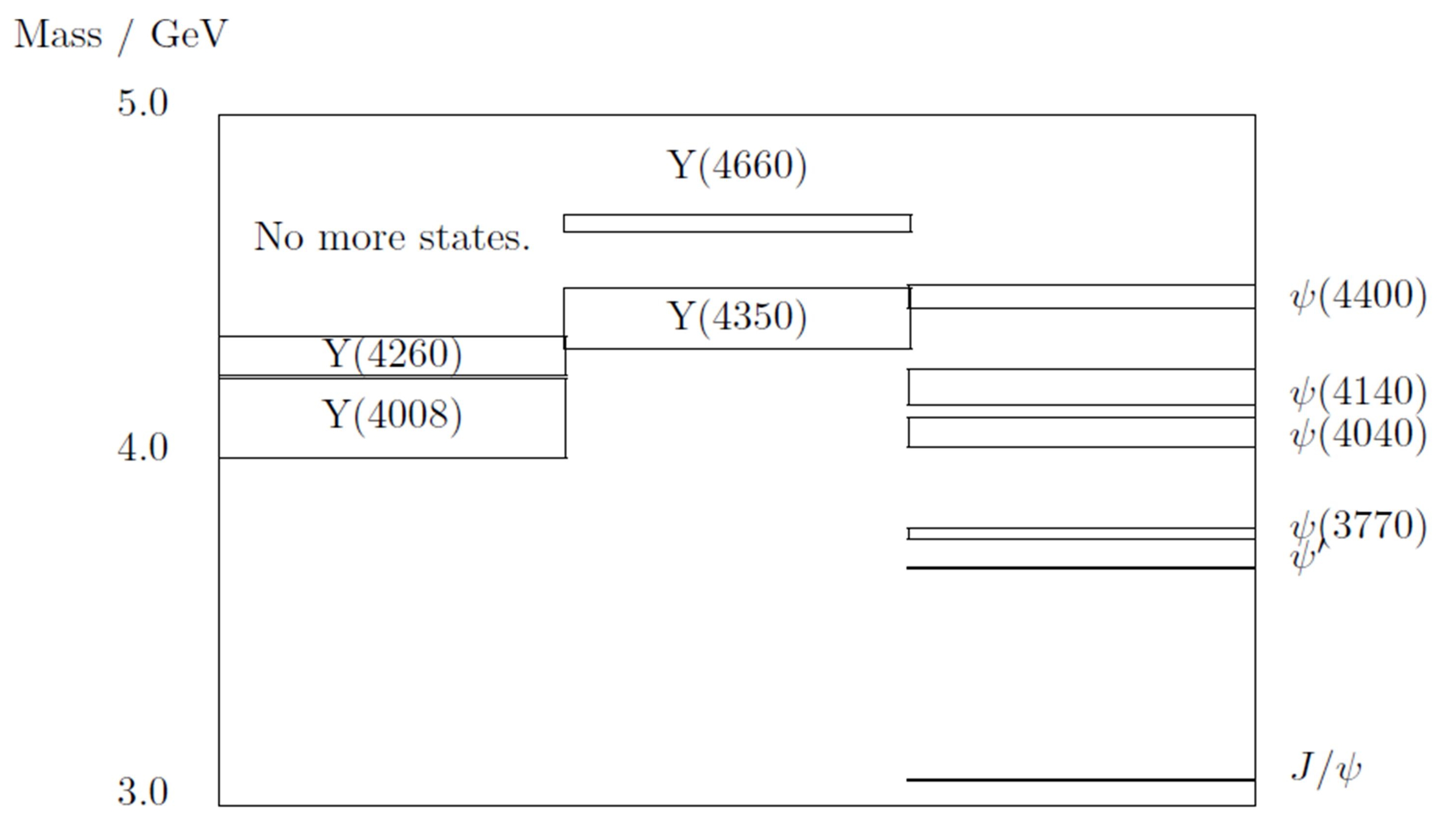}}
\caption{Level scheme for J$^{PC}$=1$^{--}$ states:
states decaying into $J$/$\psi$$\pi^+$$\pi^-$ (left column),
states decaying into $\psi'$$\pi^+$$\pi^-$ (center column),
and known $\psi$ states (radial quantum number $n$=1,...,6).\label{fy4260_scheme}}
\end{figure}

Another new charmonium-like state was observed by \BaBar
and confirmed by several experiments (see Tab.~\ref{ty4260}
for a list of the measured masses and widths) 
at a high mass of $m$$\simeq$4260~MeV,
far above the $D$$\overline{D}$ threshold. The width
is $\leq$100~MeV, which is quite narrow for such a 
high state. The observed decay is again a $\pi^+$$\pi^-$
transition to the $J$/$\psi$, similar to the above mentioned 
decay of the X(3872). However, the production mechanism 
is not $B$ meson decay but instead ISR (initial state radiation), 
i.e.\ $e^+$$e^-$$\rightarrow$$\gamma_{ISR}$Y(4260), 
i.e.\ a photon is radiated by either the $e^+$ or the $e^-$ in the initial
state, lowering the $\sqrt{s}$ and producing the Y(4260) 
by a virtual photon. 
In fact, not only one state, but four states have been observed and are shown
in Fig.~\ref{fy4260}, i.e.\ the Y(4008), the Y(4260), the Y(4250) and the Y(4660).
In a search by Belle no additional state up to $m$$\leq$7~GeV was found.
All the Y states must have the quantum numbers
$J^{PC}$=$1^{--}$, due to the observation in an initial state radiation process.
As an intriguing fact, there are known and assigned 
$J^P$=$1^{--}$ charmonium states:
$J$/$\psi$, $\psi(2S)$, $\psi$(4040), $\psi$(4160) and $\psi$(4415).
Thus, there is a clear over-population of 1$^{--}$ states 
in the $m$$\geq$4~GeV region. 
Despite partial overlap, 
apparently there seems to be no mixing: {\it (a)} no mixing among them, 
i.e.\ the Y(4008) and the Y(4260) decay to $J$/$\psi$$\pi^+$$\pi^-$, 
and the Y(4350) and the Y(4660) decay to $\psi'$$\pi^+$$\pi^-$, 
and neither of one has been observed in the other channel, and 
{\it (b)} no mixing with $\psi$ states with the Y states was observed
so far. 
The pattern of the Y states appears non-trivial (see Fig.~\ref{fy4260_scheme}): 
two non-mixing doublets without parity flip and without charge flip. 
It remains completely unclear what the underlying symmetry is.
In addition, there is no obvious pattern so far, 
how the masses of the $\psi$ states 
and the masses of the Y states might be related.

Due to their high masses, the Y states have been discussed
as possible hybrid states \cite{y4260_close_page}.
In fact, the lowest lying $[$$c$$g$$\overline{c}$$]$ $J^P$=1$^{--}$ state
was predicted by lattice QCD to have a mass $m$$\simeq$4.3~GeV \cite{y4260_bernard_meiluo}.
The interpretation as a hybrid is supported by the fact, 
that the decay Y(4260)$\rightarrow$$e^+$$e^-$ has not been observed yet. 
However, it should be allowed, as $J^{PC}$=$1^{--}$ allows coupling to a virtual photon
and subsequent $\gamma^*$$\rightarrow$$e^+$$e^-$.
\BaBar determined a very small partial decay width
$\Gamma$(Y(4260)$\rightarrow$$J$/$\psi$$\pi^+$$\pi^-$)$\times$$\Gamma$($e^+$$e^-$)/$\Gamma_{total}$
= (7.5$\pm$0.9$\pm$0.8)~eV \cite{y4260_babar_2}.
This should be compared to e.g.\
$\Gamma$($\psi'$$\rightarrow$$J$/$\psi$$\pi^+$$\pi^-$)$\times$$\Gamma$($e^+$$e^-$)/$\Gamma_{total}$
= (789$\pm$15)~eV \cite{pdg}, which is a factor $\simeq$10$^2$ higher.
A possible reason in the hybrid interpretation is,
that the decay may be blocked by the valence gluon.

\subsection{The X(4630) state}

\begin{figure}[htb]
\centerline{\includegraphics[width=0.6\textwidth]{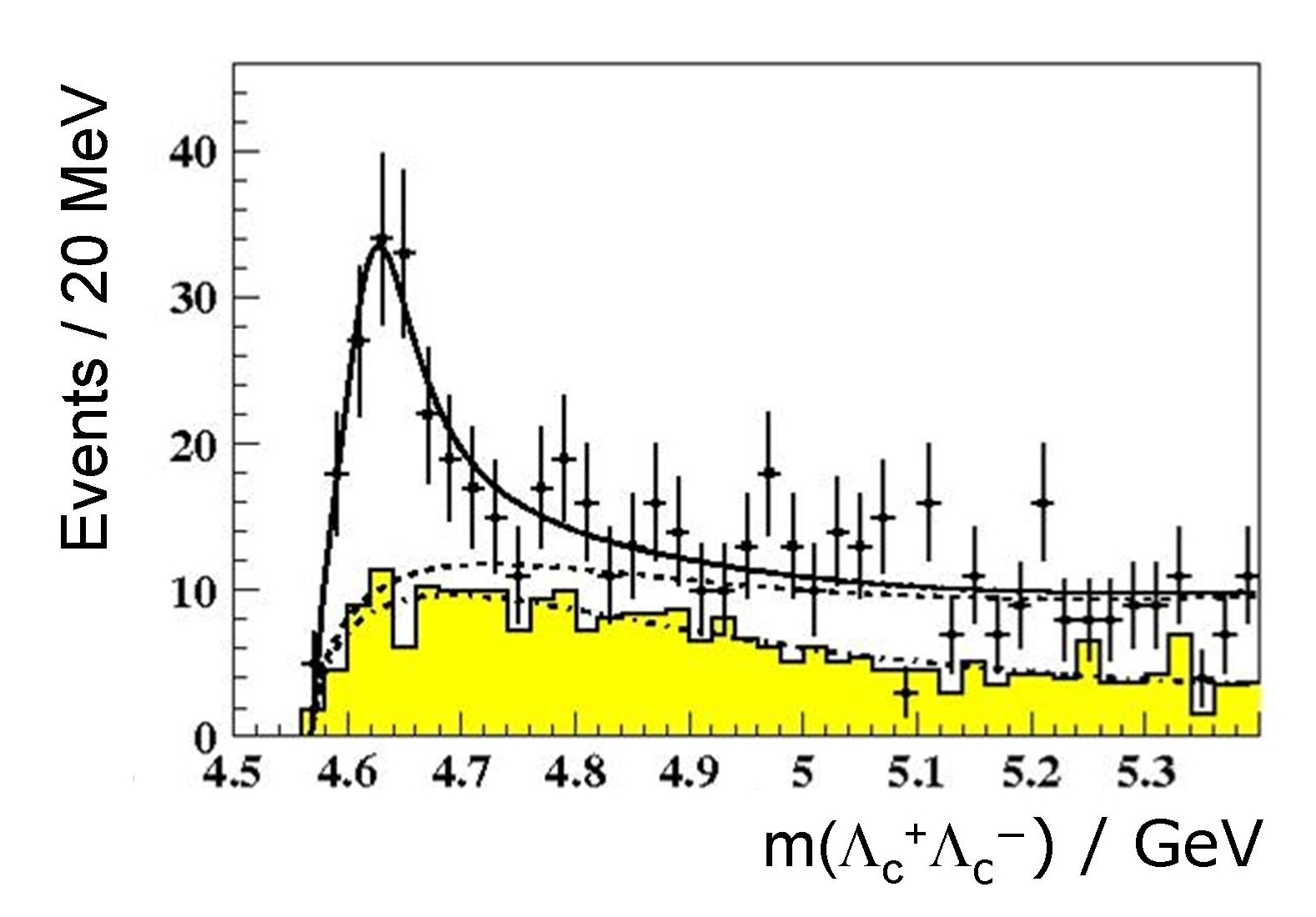}}
\caption{Invariant mass $m$($\Lambda_c^+$$\Lambda_c^-$)
for the process
$e^+$$e^-$$\rightarrow$$\gamma_{ISR}$$\Lambda_c^+$$\Lambda_c^-$
at Belle \cite{x4630_belle} showing the signal for the X(4630).
\label{fx4630}}
\end{figure}

A state which is probably identical to the Y(4660)
has also been observed at Belle \cite{x4630_belle} in the ISR production process
with a data set of 670~$fb^{-1}$,
but in the different decay channel, i.e.\ the signal was observed in
$e^+$$e^-$$\rightarrow$$\gamma_{ISR}$$\Lambda_c^+$$\Lambda_c^-$.
The state is usually refered to as X(4630).
The $\Lambda_c^+$ is reconstructed in the final states
$p$$K^0_s$($\rightarrow$$\pi^+$$\pi^-$)
$p$$K^-$$\pi^+$ and
$\Lambda$($\rightarrow$$p$$\pi^-$)$\pi^+$.
For the $\Lambda_c^-$ only partial reconstruction is used:
The recoil mass to [$\Lambda_c^+$$\gamma$] is investigated while requiring
an anti-proton (from the $\Lambda_c^-$ decay) as a tag and then a cut
around the $\Lambda_c^-$ mass is applied.
The measured mass is $m$=4634$^{+8}_{-7}$$^{+5}_{-8}$~Mev and
the measured width $\Gamma$=92$^{+40}_{-24}$$^{+10}_{-21}$~MeV.
Fig.~\ref{fx4630} shows the invariant mass
$m$($\Lambda_c^+$$\Lambda_c^-$).
A signal with a statistical significance
of 8.2$\sigma$ is observed.
The observation of this state is remarkable because of
two reasons:

\begin{itemize}

\item It is the highest charmonium state observed so far
(together with the Y(4660) of almost same mass,
but decaying into $J$/$\psi$$\pi^+$$\pi^-$), and

\item the only new state so far observed
to decay into baryons.

\end{itemize}

The potential model has an important boundary condition
for the radial wave function, which is called the
{turning point} $R_{tp}$ and can be calculated as

\begin{equation}
R_{tp} =
\frac{E-2m}{2\sigma} +
\sqrt{
\frac{4m^2-4mE+E^2}{4\sigma^2} +
\frac{4\alpha_S}{3\sigma}
}
\end{equation}

using $\sigma$=$\hbar$$c$$k$ with the string constant $k$.
This is the radius, at which {\it (a)} the Wronski
determinant must be zero and {\it (b)} the radial
wave function changes into an asymtotic, exponential tail.
For a box potential, the turning point would be
identical to $r_{box}$, and the exponential tail
of the wave function would be outside the box.
Fig.~\ref{fturning_point} shows the turning point radius
as a function of the mass.
For the X(4630), if it is a charmonium state,
the turning point is at $r_{turning point}$$>$2.1~fm.
However, a radius of $r$$\simeq$1.25~fm marks the QCD string breaking
regime. Thus, if the Y(4660) or the X(4630) are
charmonium states, it is unclear, how such a large
part of the wave function of a bound state can be
in the string breaking regime.
In any case, if it is a charmonium state,
the radial quantum number must be $n$$\geq$4.

\begin{figure}[htb]
\centerline{\includegraphics[width=0.48\textwidth]{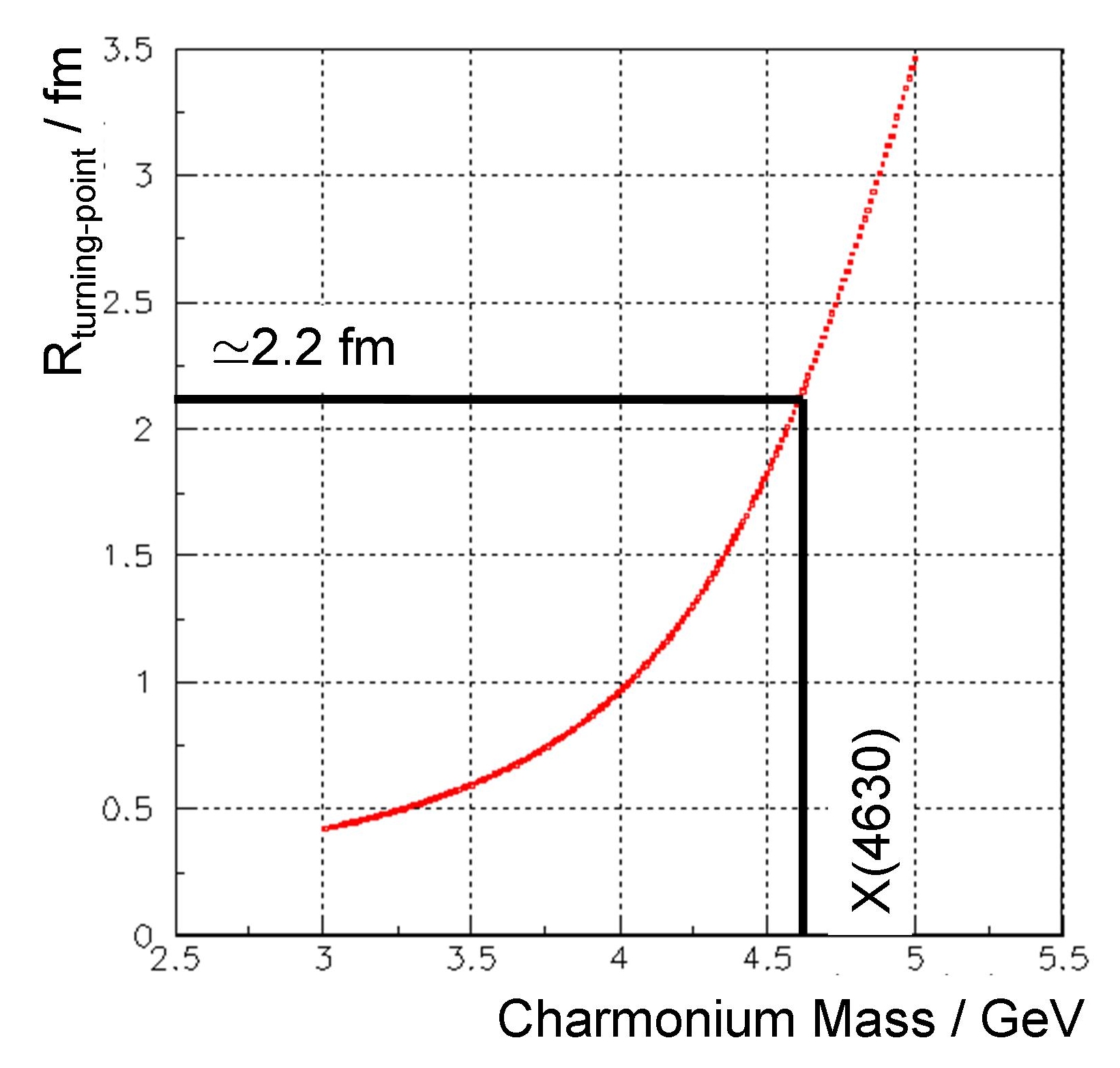}}
\vspace*{8pt}
\caption{Radius of the turning point of a $c$$\overline{c}$ wave function in the Cornell
potential vs.\ the charmonium mass.\label{fturning_point}}
\end{figure}

\subsection{The Z$_c$(3900) state}

A new state, tentatively called the Z$_c^+$(3900), was observed by
BESIII \cite{zc3900_bes3} in the decay of the Y(4260)$\rightarrow$Z$_c^+$(3900)$\pi^{\pm}$
in a data set of 525~$pb^{-1}$. 
BESIII is operating with center-of-mass energies in the charmonium mass region, 
and producing the Y(4260) directly via $e^+$$e^-$$\rightarrow$Y(4260) at $\sqrt{s}$=4.26~GeV.
Importantly the Z$_c^+$(3900) is a charged state, and thus can not be a charmonium state.
A charged combination could be formed by a state composed of four quarks.
This may be tetraquark state (such as $[$$c$$u$$\overline{c}$$\overline{d}$$]$)
or a molecular state (such as $D^\pm$$\overline{D}^{0*}$). 
The Z$_c^+$(3900) was reconstructed in the decay to $J$/$\psi$$\pi^{\mp}$.
Fig.~\ref{fzc3900} (left) shows the observed signal, which has a statistical
significance of $>$8$\sigma$. From the two charged pions, the one is used,
which gives the higher invariant mass for $J$/$\psi$$\pi^{\pm}$,
in order to remove combinatorical background from the charged pion
of the Y(4260) transition to the new state.
The measured mass is $m$=3899.0$\pm$3.6$\pm$4.9~MeV and the measured
width $\Gamma$=46$\pm$10$\pm$20~MeV.
The observation of this new state is remarkable because this state 
seems to provide 
for the first time a connection between the Z states and the Y states,
possibly pointing to the same interpretation of their nature.
Only a few days later, the state was confirmed by Belle \cite{zc3900_belle} in the same decay channel
$J$/$\psi$$\pi^{\mp}$ and also in Y(4260) decays, while in the Belle case the Y(4260) was produced
in the ISR process $\Upsilon$($n$$S$)$\rightarrow$$\gamma_{ISR}$Y(4260).
The measure mass of $m$=3894.5$\pm$6.6$\pm$4.5~MeV and width $\Gamma$=63$\pm$24$\pm$26~MeV are
both consistent with the BESIII measurement.
Fig.~\ref{fzc3900} (right) shows the observed signal, which has a statistical
significance of $>$8$\sigma$ in a data set of 967~$fb^-1$.
Again, as the Z$_c^+$(4430), the state was observed as Z$_c^+$(3900) and Z$_c^-$(3900)
with about the same yield \cite{zc3900_bes3}, indicating a doublet.
Concerning the quantum numbers, remarkably the isospin must be $I$=1,
(as the isospin of the pion is $I$=1), if we assume $I$=0 for the Y(4260).
If the heavy meson pair is assumed to be in the $S$-wave,
the spin-parity of the state is uniquely determined as $J^P$=$1^+$.
$C$-parity $(-1)^{L+S}$ is only defined for neutral particles,
thus there can only be a $G$-parity assignment to the Z$_c^+$(3900).
The $G$-parity $(-1)^{L+S+I}$ with $L$=0, $S$=1 and $I$=1 thus gives $G$=+.
As $G$-parity should be preserved in strong decays,
this assignment, due to the $G$-parity $G$=$-$ for the pion, has the
interesting implication that the Y(4260) would have $G$=$-$.
This would be compatible with an $I$=0 isosinglet assignment for the Y(4260),
which has the important implication, that there is no charged partner 
of the Y(4260) existing.

\begin{figure}[htb]
\centerline{\includegraphics[width=1.0\textwidth]{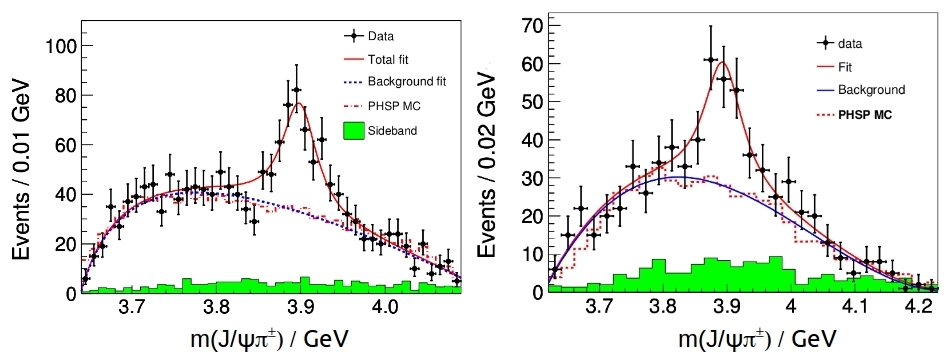}}
\caption{$J$/$\psi$$\pi^{\mp}$ invariant mass in Y(4260) decays,
indicating the Z$_c^+$(3900) signal for BESIII (left) \cite{zc3900_bes3}
and Belle (right) \cite{zc3900_belle}. For details see text.
\label{fzc3900}}
\end{figure}

\subsection{A $D$-wave state}

\label{D-wave}

\begin{figure}[htb]
\centerline{\includegraphics[width=0.6\textwidth]{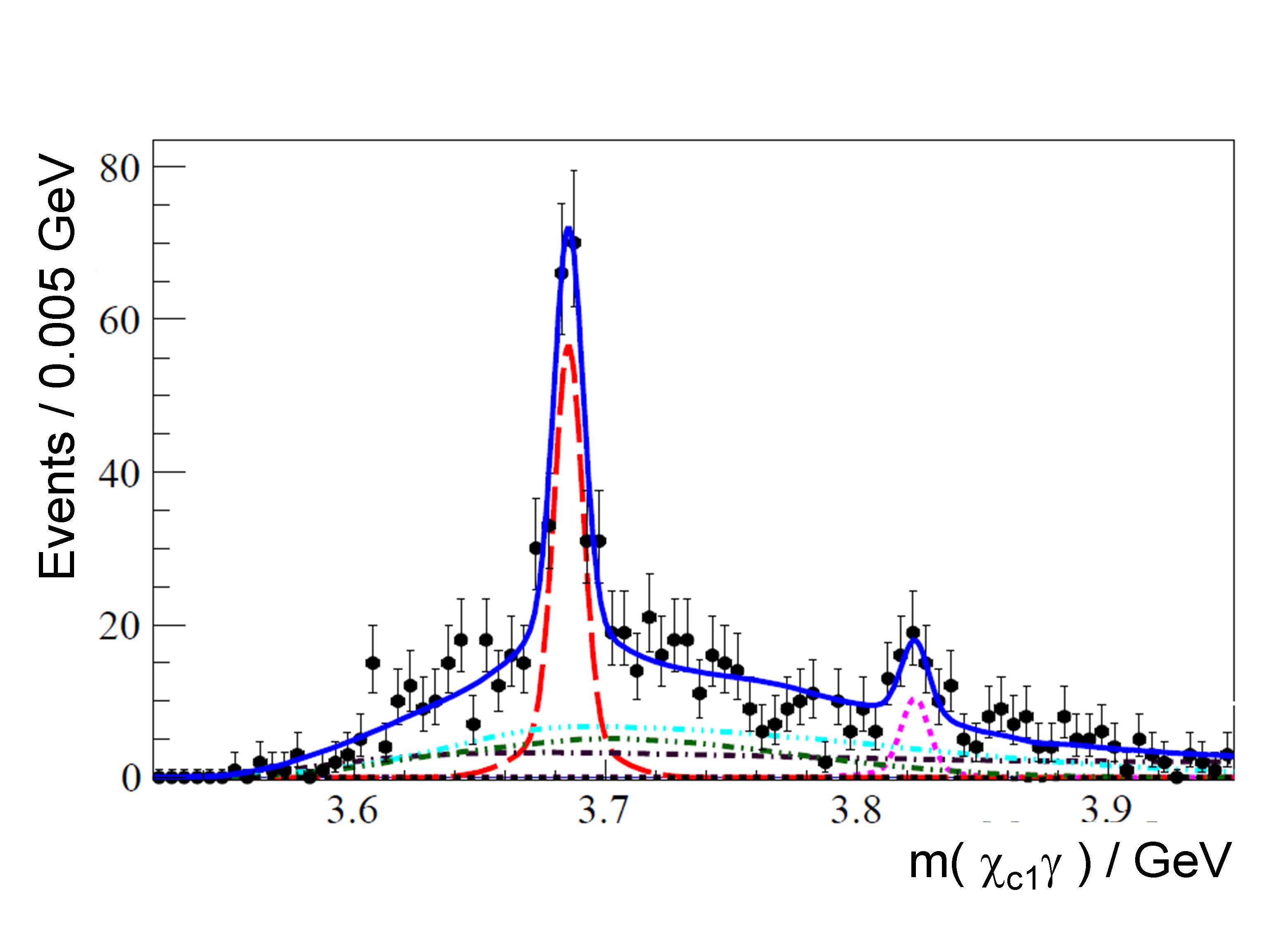}}
\vspace*{8pt}
\caption{Invariant mass $m$($\chi_{c1}$$\gamma$) in $B$ meson decays at Belle \cite{D-wave_belle}
showing the signal of the $^3D_2$ charmonium candidate X(3820). The dashed, dotted 
and dash-dotted line represent different backgrounds (combinatorial, peaking and non-peaking 
background from $\psi'$ and X(3872) decays other than $\chi_{cJ}$$\gamma$, respectively,
where the term peaking refers to peaking in $m_{BC}$).\label{fdwave}}
\end{figure}

Belle investigated the decay
$B^+$$\rightarrow$$K^+$$\chi_{c1}$$\gamma$ with 
$\chi_{c1}$$\rightarrow$$J$/$\psi$$\gamma$
using a data set of 711~$fb^{-1}$ \cite{D-wave_belle}.
A search for charmonium(-like) states decaying to $\chi_{c1}$$\gamma$
was performed. Fig.~\ref{fdwave} shows the invariant mass 
$m$($\chi_{c1}$$\gamma$).
In other words, the search was based upon a sequence 
of two radiative decays with both $\Delta$$L$=1. 
The radiative transition also flips the parity
due to the quantum numbers of the photon $J^P$=1$^-$,
and therefore the requirement of the intermediate 
$\chi_{c1}$ with positive parity. 
A new state at a mass of 3823.1$\pm$1.8$\pm$0.7~MeV
was observed with a 3.8$\sigma$ significance. 
The product branching fraction was measured 
as ${\cal B}$($B^+$$\rightarrow$$K^+$X(3820))$\times$
${\cal B}$(X(3820)$\rightarrow$$\chi_{c1}$$\gamma$)=
(9.7$^{+2.8}_{−2.5}$$^{+1.1}_{−1.0})$$\times$10$^{−4}$,
which is a factor $\simeq$10 larger than e.g.\ 
the sum of all measured product branching fractions
of the X(3872). 
The observed state might be one of the charmonium
$D$-wave ($L$=2) states, as such states should primarily decay radiatively
to $\chi_{cJ}$ states by $L$=2$\rightarrow$$L$=1 
transitions and according branching fractions
should be high $\geq$50\% \cite{quigg_D-wave} \cite{eichten_lane_quigg}.
There are four expected $n$=1 $D$-wave states:
the $\eta_{c2}$ ($^1D_2$) 
with $J^{PC}$=2$^{-+}$ 
and $\psi_{1,2,3}$ ($^3D_{1,2,3}$) 
with $J^{PC}$=1,2,3$^{--}$.
The prediction \cite{potential_2005} for the 
$\psi_1$ ($^3$$D_1$) of 3.7699~GeV is much lower
than the observed X(3820). 
The $\psi_3$ ($^3$$D_3$) can not decay radiatively
by an E1 transition and should thus be suppressed.
The $\eta_{c2}$ ($^1D_2$) would require a spin-flip
in the transition, and should be suppressed as well.
The only candidate, which fulfills all the required
properties, is the $\psi_2$ ($^3$$D_2$) state 
with $J^{PC}$=2$^{--}$ and a predicted mass
3.838~MeV \cite{potential_2005}, 
which is close to the observed mass.
In addition, the $\psi_2$ is predicted to be narrow
$\Gamma$$\simeq$300-400~keV \cite{quigg_D-wave}, consistent with a 
preliminary measured width $\Gamma$=4$\pm$6~MeV.
As the observed state is above the open charm 
thresholds (3730~MeV for $D^0$$\overline{D}^0$ and 
3739~MeV for $D^+$$D^-$, respectively), decays into final states
with charm should be expected. However, for the $\psi_2$
the decay 2$^{--}$$\rightarrow$$0^{-+}$0$^{-+}$ 
with $\Delta$$L$=2 (i.e.\ ($-$1)$^L$=+1) 
is forbidden by parity conservation, and thus other decays (such as
the observed one) should be enhanced. 
This mechanism could explain the high observed branching fraction
into $J$/$\psi$$\gamma$$\gamma$.
Note that the decays to $D$$\overline{D}^*$ or $D^*$$\overline{D}^*$
are forbidden by energy conservation.
The observed product branching fraction is consistent
with a calculation with color-octet amplitudes \cite{D-wave_theory}
predicting 
${\cal B}$($B$$\rightarrow$$K$$^3D_2$)$\times$
${\cal B}$($^3D_2$$\rightarrow$$\chi_{c1}$$\gamma$)=
(3.7$-$7.5)$\times$10$^{-4}$.

\section{Bottomonium(-like) states}

\subsection{The $h_b$(1P) and the $h_b$(2P)}

\begin{figure}[hhh]
\centerline{\includegraphics[width=0.8\textwidth]{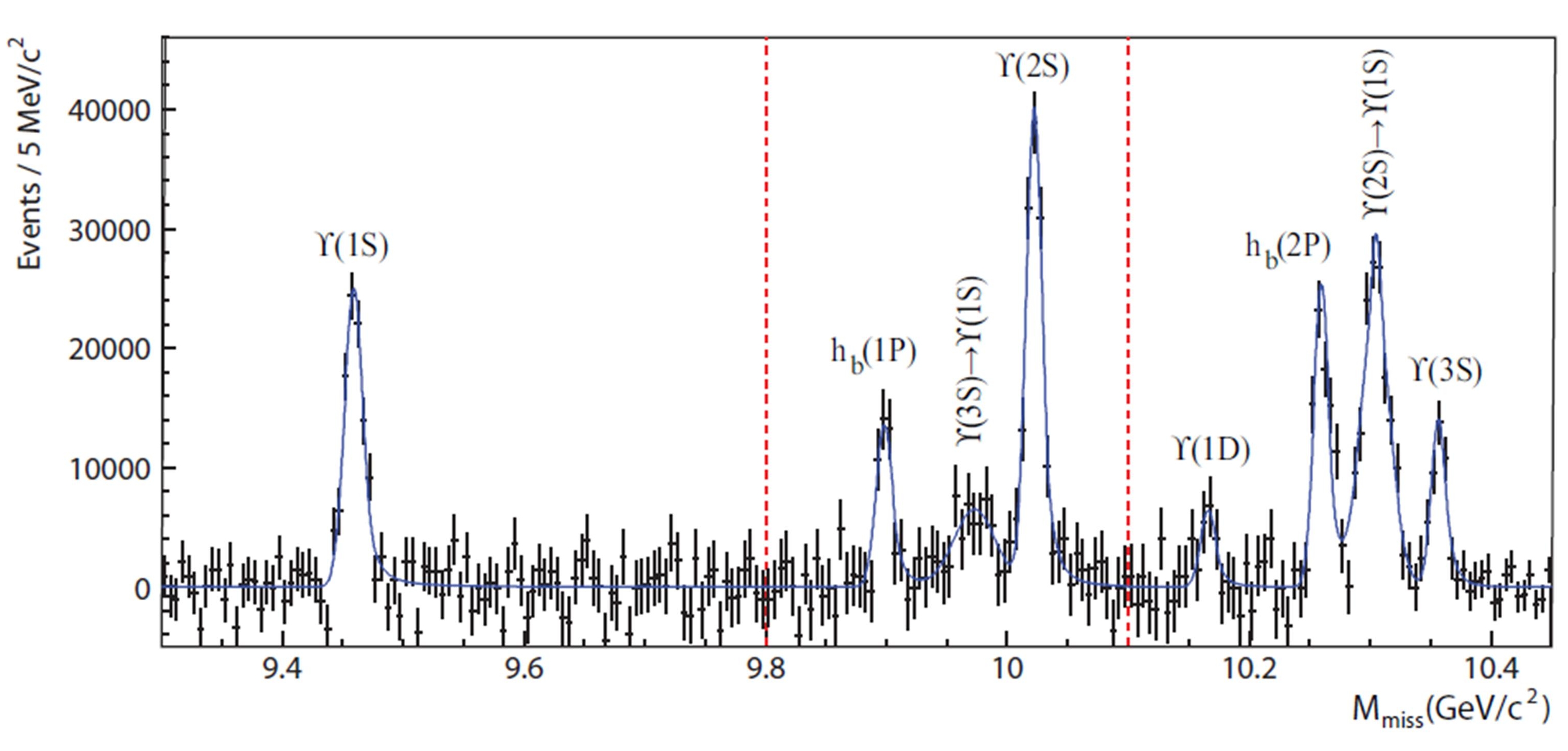}}
\caption{Observation of the $h_b$(1S) and $h_b$(2S) at Belle. 
For details see text.\label{fhb}}
\end{figure}

\begin{figure}[hhh]
\centerline{\includegraphics[width=0.8\textwidth]{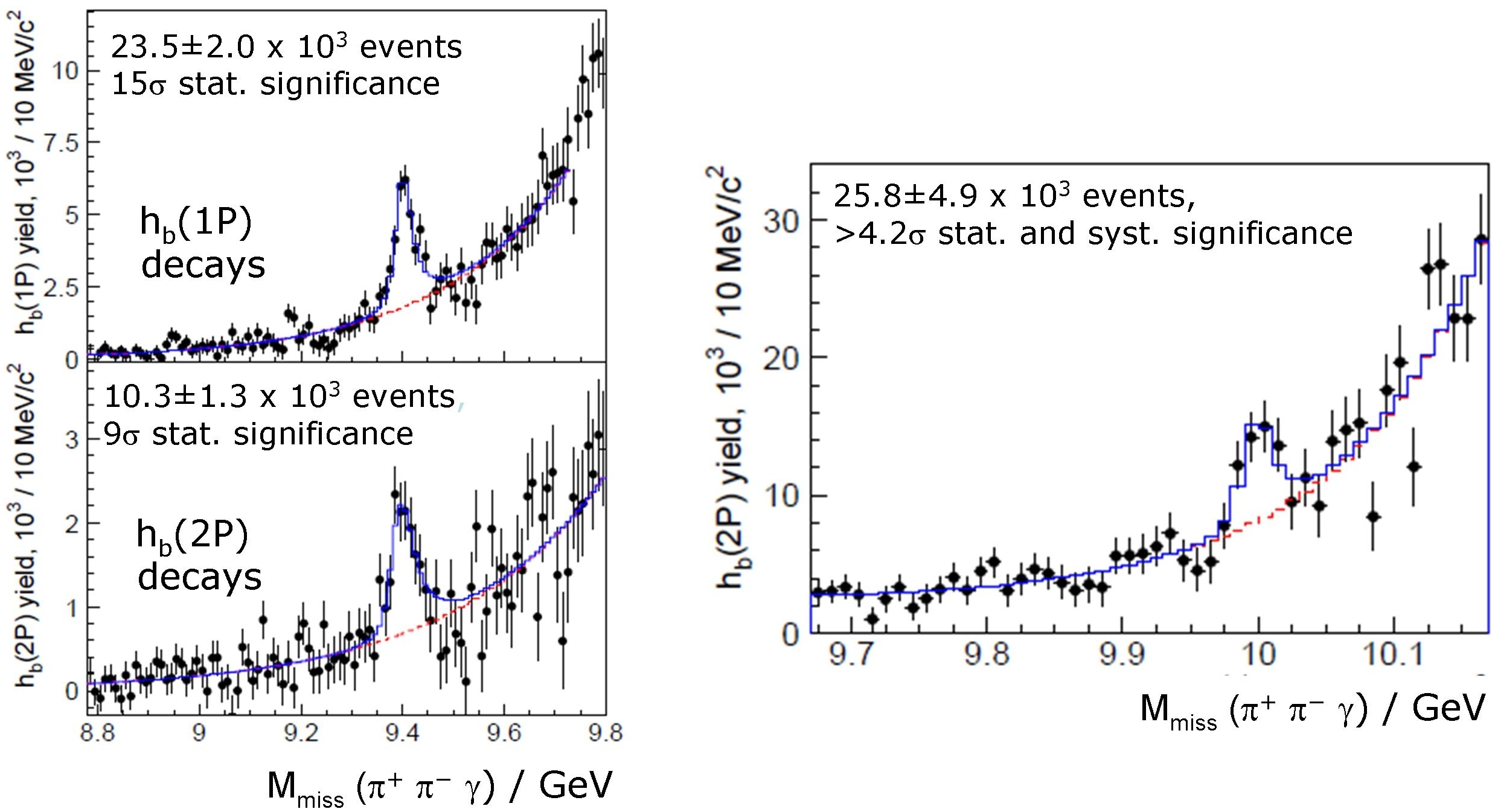}}
\caption{Observations of the $\eta_b$(1$P$) (left) and $\eta_b$(2$P$) (right) 
at Belle. For details see text.\label{fetab}}
\end{figure}

\noindent
In a recent analysis by Belle, a particular technique was used, 
namely the study of {\it missing mass} to
a $\pi^+$$\pi^-$ pairs in $\Upsilon$(5S) decays \cite{hb_belle}.
Fig.~\ref{fhb} shows the background-subtracted 
missing mass for a $\Upsilon$(5S) data set of 121.4~fb$^{-1}$. 
Among several known states such as the $\Upsilon$(1S), $\Upsilon$(2S), 
$\Upsilon$(3S) and $\Upsilon$(1D), 
there are addititional peaks arising from the transistions 
$\Upsilon$(3S)$\rightarrow$$\Upsilon$(1S)$\pi^+$$\pi^-$
$\Upsilon$(2S)$\rightarrow$$\Upsilon$(1S)$\pi^+$$\pi^-$, 
with the $\Upsilon$(3S) and $\Upsilon$(2S) 
being produced in the decay of the primary $\Upsilon$(5S). 
In addition to the expected signals, first observations 
of the bottomonium singlet 
$P$-wave states $h_b$(1P) and $h_b$(2P) were made. 
Their measured masses are 
$m$=9898.3$\pm$1.1$^{+1.0}_{-1.1}$~MeV and 
$m$=10259.8$\pm$0.6$^{+1.4}_{-1.0}$~MeV, respectively.
The red, dashed lines in Fig.~\ref{fhb} indicate regions
of different paramtrisations of the background. 
For the $h_b$, this measurement is consistent with the first evidence
(3.1$\sigma$ stat.\ significance) by BaBar in $\Upsilon$(3S) decays 
with a mass of 9902$\pm$4(stat.)$\pm$2(syst.)~MeV \cite{hb_babar_pi0}.
The masses can be compared to predictions from 
potential model calculations \cite{potential_bb}
with 9901 MeV and 10261 MeV, respectively, 
i.e.\ the deviations are only 2.7 MeV and 1.2~MeV.


\subsection{The $\eta_b$(1S) and the $\eta_b$(2S)}

The $\eta_b$(1$S$) is the bottomonium ground state 
$1^1S_0$ with J$^{PC}$=0$^{-+}$.
It was discovered by BaBar in the radiative decay 
$\Upsilon$(3$S$)$\rightarrow$$\gamma$$\eta_b$.
The measured mass was 9388.9$^{+3.1}_{-2.3}$(stat)$\pm$2.7(syst)~MeV, 
The observation was confirmed by CLEO III using 6 million Upsilon(3S) decays
with a measured mass $m$=9391.8$\pm$6.6$\pm$2.0~MeV.
The observation of the $h_b$ (see above) by Belle also enabled a search
for the radiative decay $h_b$(1$P$)$\rightarrow$$\eta_b$(1$S$)$\gamma$, 
which was observed with a very high significance $>$13$\sigma$
in a dataset of 133.4~fb$^{-1}$ at the $\Upsilon$(5$S$) and in the 
nearby continuum \cite{etab_belle}. In addition, even the $\eta_b$(2$S$)
was observed in $h_b$(2$P$)$\rightarrow$$\eta_b$(2$S$)$\gamma$.
Fig.~\ref{fetab} shows the 
$\pi^+$$\pi^-$$\gamma$ missing mass for the case of the $\eta_b$(1$S$) (left)
and $\eta_b$(2$S$) (right), 
where the charged pion pair originates from the transition
$\Upsilon$(5$S$)$\rightarrow$$h_b$(1$P$,2$P$)$\pi^+$$\pi^-$.
The measured masses are $m$($\eta_b$(1$S$))=9402.4$\pm$1.5$\pm$1.8~MeV
and $m$($\eta_b$(2$S$))=9999.0 $\pm$3.5$^{+2.8}_{-1.9}$~MeV.
Due to the high resolution, this measurement also enabled
the measurement of the width of the $\eta_b$ as 
$\Gamma$=10.8$^{+4.0}_{-3.7}$$^{+4.5}_{-2.0}$,
which is consistent with the expectation from 
potential models to 5$\leq$$\Gamma$$\leq$20~MeV.
The measurements of the $\eta_b(1S)$ and $\eta_b(2S)$ allow
precision determination of the hyperfine mass splittings 
$\Upsilon$(1$S$)-$\eta_b$(1$S$) and $\Upsilon$(2$S$)-$\eta_b$(2$S$),
using the masses of the $\Upsilon$(1$S$) and $\Upsilon$(2$S$)
from \cite{pdg}. The mass splittings are listed in Tab.~\ref{thfsplit}.
The splittings are in good agreement 
with the expectation from a potential model with relativistic 
corrections \cite{potential_bb} or lattice QCD calculations 
with kinetic terms up to {\cal O}($v^6$) 
\cite{lattice_bb_new_meinel}. 
However, lattice QCD calculations to {\cal O}($v^4$) with charm sea
quarks predict higher splittings which are $\simeq$10~MeV larger. 
Note that perturbative non-relativistic QCD calculations
up to order ($m_b$$\alpha_S$)$^5$ predict significant 
smaller splittings e.g.\ 39$\pm$11$^{+9}_{-8}$~MeV
\cite{pNRQCD_bb}. 

\begin{table}[tbh]
\begin{tabular}{|l|l|l|l|l|}
\hline
&  Belle \cite{etab_belle} & Potential \cite{potential_bb} & LQCD \cite{lattice_bb_new_hpqcd}) & LQCD \cite{lattice_bb_new_meinel}\\
\hline
$\Upsilon(1S)$$-$$\eta_b$ & 57.9$\pm$2.3 MeV & 60.0 & 70$\pm$9 MeV & 60.3$\pm$5.5$\pm$5.0$\pm$2.1~MeV\\
\hline
$\Upsilon(2S)$$-$$\eta_b'$ & $24.3^{+4.0}_{-4.5}$ MeV & 30.0 & 35$\pm$3 MeV & 23.5$\pm$4.1$\pm$2.1$\pm$0.8~MeV\\
\hline
\end{tabular}
\caption{Bottomonium hyperfine splittings: measurement, 
calculated by potential model and calculated by Lattice QCD (LQCD).\label{thfsplit}}
\end{table}

\section{Test of the tensor term in the potential}

The measured masses of the $h_b$ and $h_b'$ can be used for a precision test
of the hyperfine splitting in the Cornell-type potential (Eq.~\ref{ecornell}), 
i.e.\ a test of the relation

\begin{equation}
m(h_b) \stackrel{?}{=} \frac{ m(\chi_{b0}) + 3 \cdot m(\chi_{b1}) + 5 \cdot m(\chi_{b2})}{9} 
\end{equation}

\noindent
using the world average masses of the $\chi_{b0,1,2}$ and $\chi_{b0,1,2}'$ from \cite{pdg}.
The hyperfine splitting $\Delta$$m_{HF}$=$<$$m$($n^3P_J$)$-$$m$($n^1P_1$) was measured 
as $\Delta$$m_{HF}$=(+1.6$\pm$1.5)~MeV for $n$=1 and $\Delta$$m_{HF}$=(+0.5$^{+1.6}_{-1.2})$~MeV
for $n$=2. This can be used as a test for the tensor term in the potential

\begin{equation}
V_{tensor} = \frac{1}{m^2} \frac{4\alpha_S}{r^3} 
( \frac{3\vec{S_1}\vec{r} \cdot \vec{S_2}\vec{r}}{r^2}
- \vec{S}_1 \vec{S}_2 )
\end{equation}

with the spins of the heavy quarks $\vec{S}_1$ and $\vec{S}_2$, the heavy quark mass $m$ 
and the quark antiquark distance $r$, which is usually treated
as a perturbation in the potential. 
It vanishes for $S$=0 (e.g.\ $\eta_b$, $\Upsilon$($n$$S$), $h_b$, ...) and
$L$=0 (e.g.\ $^1D_2$ state, ...). 
In a simplified view, a non-zero $\Delta$$m_{HF}$ would mean, that the wavefunction
of the $h_b$ at $r$=0 is non-vanishing. 
The sign of the potential term is positive, thus masses should be shifted up.
Although the above mentioned measurements of $\Delta$$m_{HF}$ are consistent 
with zero, however positive values seem to be preferred for the $b$$\overline{b}$
case, mildly suggesting to indicate an effect of the tensor term.
This can be compared to measurements of
$\Delta$$m_{HF}$=0.02$\pm$0.19$\pm$0.13~MeV \cite{hc_cleo_2} and 
$\Delta$$m_{HF}$=−0.10$\pm$0.13$\pm$0.18 MeV \cite{hc_bes3}
charmonium system, (i.e.\ the $h_c$).

\subsection{Test of flavor independence of the potential}

The new mass measurements in the bottomonium region enable for the first time a precision test 
of the flavour independance of the $c$$\overline{c}$ and $b$$\overline{b}$ 
systems. The important question is, if the level spacing is independant from the 
quark mass. According to \cite{cc_bb_quigg_test}, for a potential of
the form $V$($r$)=$\lambda$$r^\nu$ the level spacing is 
$\Delta$$E$$\propto$(2$\mu$/$\hbar^2$)$^{-\nu/(2+\nu)}$$|$$\lambda$$|$$^{2/(2+\nu)}$.
where $\mu$ is the (reduced) quark mass. For a pure Coulomb potential
($\nu$=$-$1), which should be dominating for the low lying states, 
this leads to $\Delta$$E$$\propto$$\mu$, This would imply that 
the level spacing would increase linearly with mass, i.e.\
$\Delta$$E$($b$$\overline{b}$)$\simeq$3$\Delta$$E$($c$$\overline{c}$).
For a pure linear potential it would be 
$\Delta$$E$$\propto$$\mu^{-1/3}$, thus the level spacing would 
decrease for higher quark masses, i.e.\ 
$\Delta$$E$($b$$\overline{b}$)$\simeq$0.5$\Delta$$E$($c$$\overline{c}$).
As can be seen in Fig.~\ref{fcc_bb_quigg_test}, 
for the mass splittings involving the $h_b$ (S=0, L=1) 
the agreement between
$c$$\overline{c}$ and $b$$\overline{b}$ is excellent,
i.e.\ 10.2 vs.\ 10.1~MeV and 43.9 vs.\ 43.8~MeV.
There are two possible explanations of this
remarkable symmetry. 
{\it (1)} For a pure logarithmic potential V(r)=$\lambda$ln$r$ (i.e.\ the limit
$\nu$$\rightarrow$0) the level spacing is 
$\Delta$$E$$\propto$$\lambda$$\mu^0$. This means, 
the flavour independance would be 
strictly fulfilled. 
{\it (2)} The other way to reach the flavour independance is,
that the Coulomb potential with 
$\Delta$$E$($b$$\overline{b}$)$\simeq$3$\Delta$$E$($c$$\overline{c}$)
(see above) and the linear potential with 
$\Delta$$E$($b$$\overline{b}$)$\simeq$0.5$\Delta$$E$($c$$\overline{c}$)
(see above) cancel each other quantitatively in an exact way. 
It also implies that the size of the according $\lambda$ pre-factors 
($\lambda$=$-$4/3$\alpha_S$ for the Coulomb-like potential and 
$\lambda$=$k$ for the linear potential) just seem to have
the exactly correct size assigned by nature in a fundamental way.\\
For the ground states ($S$=0, $L$=0) the agreement
of the mass splittings between $c$$\overline{c}$ and $b$$\overline{b}$
is not as good, i.e.\ 65.7 vs.\ 59.7~MeV, and may point to the fact, that
there is an additional effect which lowers the $\eta_c$ mass. 
This might be mixing of the $\eta_c$ with the light quark states 
of the same quantum number 0$^{-+}$ (i.e.\ $\eta$ or $\eta'$). 

\begin{figure}[tbh]
\centerline{\includegraphics[width=0.8\textwidth]{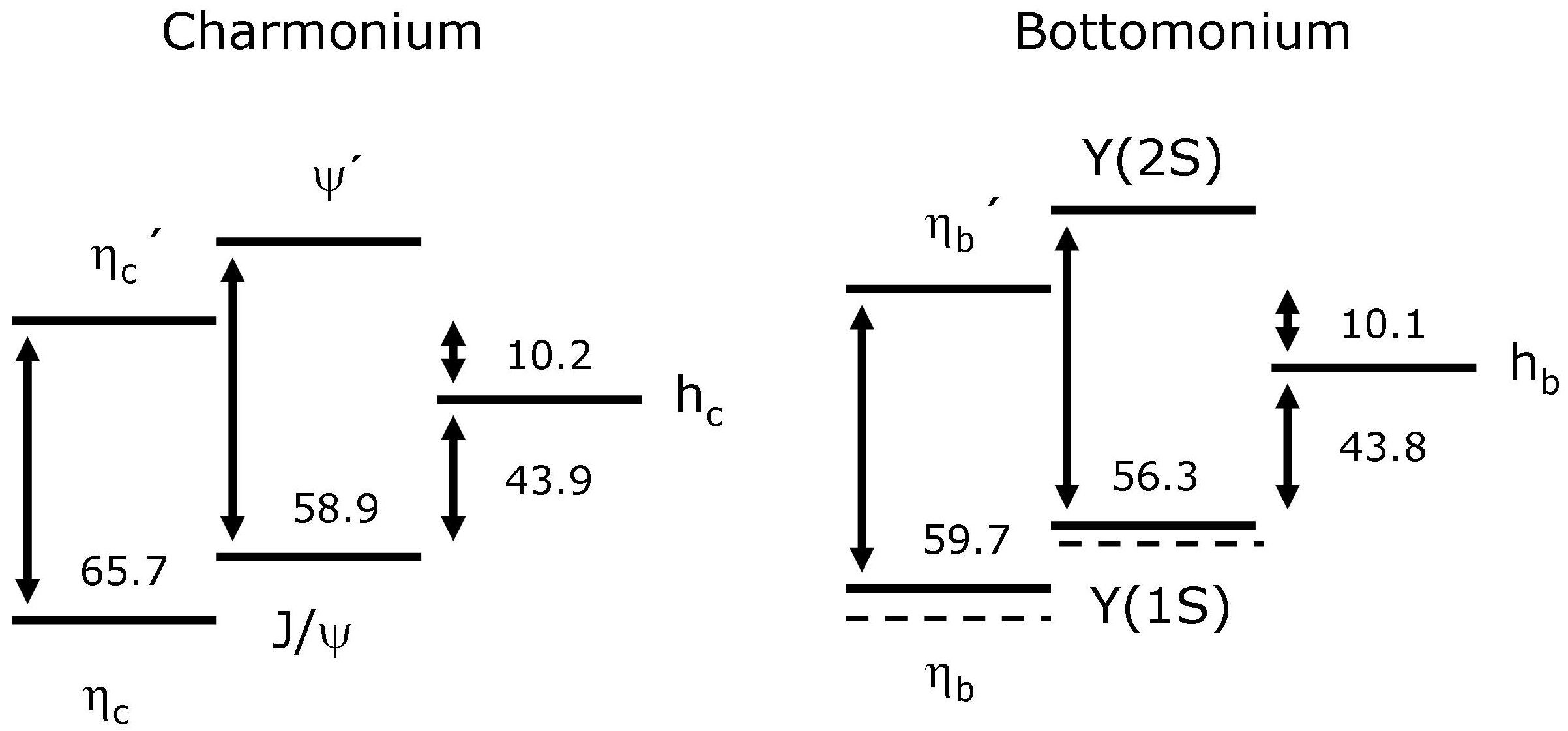}}
\caption{Mass splittings (in MeV) based upon the new measurements \cite{etab_belle} 
of the $h_b$, $\eta_b$ and $\eta_b'$, 
using masses from \cite{pdg} for the other states, for charmonium (left) and bottomonium (right).
The dotted lines indicate levels for the theoretical case 
of exact flavour independance.\label{fcc_bb_quigg_test}}
\end{figure}

\subsection{The Y$_b$(10889) state}

While investigating $\Upsilon$(5$S$) decays, Belle discovered a highly anomalous behavior.
For the $\Upsilon$(5$S$), the beam energies of the KEK-B accelerator were changed in a way,
to keep the center-of-mass boost the same as on the $\Upsilon$(4$S$) resonance. 
Thus, all analysis techniques could be applied.
In a data set of 21.7~$fb^{-1}$, the processes $e^+$$e^-$$\rightarrow$$\Upsilon$($n$$S$)$\pi^+$$\pi^-$ 
with $n$=1,2,3 were investigated. 

First of all, the cross section of decays to the Y(1$S$) was found to be anomalously large.
While in a data set of 477~$fb^{-1}$ on the $\Upsilon$(4$S$), 
$N$=44$\pm$8 events of $\Upsilon$(1$S$)$\pi^+$$\pi^-$ 
were observed \cite{belle_4Sto1Spipi}, in the data set of 21.7~$fb^{-1}$ on the $\Upsilon$(5$S$) 
$N$=325$\pm$20 events of $\Upsilon$(1$S$)$\pi^+$$\pi^-$ were observed \cite{belle_5Sto1Spipi}.
This means, that in a data set corresponding to $\simeq$1/20 the size of the data set
and $\simeq$1/10 of the production cross section, still a factor 7.4 more events are observed.
This corresponds in total to a signal, which is more than a factor 10$^3$ higher than
the expectation.
In addition, not only the $\Upsilon$(5$S$)$\rightarrow$$\Upsilon$(1$S$)$\pi^+$$\pi^-$ 
but also the $\Upsilon$(5$S$)$\rightarrow$$\Upsilon$(2$S$)$\pi^+$$\pi^-$ 
was found to be larger than expected by more than a factor 5$\times$10$^2$.
Note that $\Upsilon$(5$S$)$\rightarrow$$\Upsilon$(4$S$)$\pi^+$$\pi^-$ is kinematically suppressed.

One of the possible explanation for the observed anomalously high yield 
was a new resonance nearby the $\Upsilon$(5$S$), decaying into the same final state \cite{yb_hou}.
Therefore a beam energy scan was performed \cite{belle_yb}.
Typical step sizes in the variation of the $\sqrt{s}$ were 6-10 MeV.
On each scan point more than 30~pb$^{-1}$ were performed.
For each energy point, the yield of $\Upsilon$(1$S$,2$S$,3$S$)$\pi^+$$\pi^-$ 
was determined by an unbinned maximum likelihood fit. 

Fig.~\ref{fyb_scan} (bottom) shows the fitted signal yield as a function of $\sqrt{s}$.
These excitation curves are fitted with a 
Breit-Wigner shape with floating mean and width,
but constraint to be identical parameters for all three curves.
The normalizations for the three curves are floating independently.
The fitted mean is at $\simeq$20~MeV higher mass 
and the width is about a factor $\simeq$2 narrower than the $\Upsilon$(5$S$).
This indicates that the observed resonance is not the $\Upsilon$(5$S$), 
but instead a new state which was given the name Y$_b$(10889).

For comparison, Fig.~\ref{fyb_scan} (top) shows the ratio $R_b$ vs.\ $\sqrt{s}$,
where $R_b$ is defined as the ratio of the inclusive hadronic cross section 
$\sigma$($e^+$$e^-$$\rightarrow$hadrons) to $\sigma$($e^+$$e^-$$\rightarrow$$\mu^+$$\mu^-$).
The final measurement for the new state yields a mass of $m$=10888.4$^{+2.7}_{−2.6}$$\pm$1.2~MeV
and a width of $\Gamma$=30.7$^{8.3}_{7.0}$$\pm$3.1~MeV.
The final results for the widths, as measured in the resonance scan, 
are summarized in Tab.~\ref{tyb10889}

As the Y$_b$(10889) does not coincide with a threshold, 
it cannot be interpreted as a molecule, neither as a threshold effect.
there must be another explanation for its nature. 
The lowest lying tetraquark state with $J^{PC}$=1$^{--}$ is predicted
at a mass $m$=10.890~MeV \cite{yb_tetraquark}, 
well consistent with the experimental observation. 
It would be a $[$$bq$$\overline{b}$$\overline{q}$$]$ tetraquark,
where $q$ denotes a light $u$ or $d$ quark, which are assumed 
to have the identical consituent mass of 305~MeV. 
In addition, the tetraquark model could explain the observed anomalous yield \cite{yb_tetraquark}, . 
If the Y$_b$(10889) is a pure $b$$\overline{b}$ state,
there are no light quarks in the initial state.
The $\pi^+$$\pi^-$ pair in the 
$\Upsilon$(5$S$)$\rightarrow$$\Upsilon$(1$S$,2$S$,3$S$)$\pi^+$$\pi^-$
transition must be created by two gluons and subsequent 
$g$$\rightarrow$$u$$\overline{u}$, $g$$\rightarrow$$d$$\overline{d}$,
and rearrangement to $u$$\overline{d}$ and $d$$\overline{u}$. 
Thus, the transition would be Zweig forbidden.
If the Y$_b$(10889) is a $[$$bq$$]$$[$$\overline{b}$$\overline{q}$$]$ tetraquark,
then there is a $u$$\overline{u}$ or $d$$\overline{d}$ already present
in the initial state, and only one additional pair must be 
formed from the QCD vacuum. Thus, the transition is Zweig allowed
and the transition rate would be increased.
An effect, which could explain the observed properties of the Y$_b$(10889),
however without assuming an exotic nature, is {\it rescattering}. 
In the rescattering model, 
the decay $\Upsilon$(5$S$)$\rightarrow$$\Upsilon$(1$S$,2$S$,3$S$)$\pi^+$$\pi^-$
would not proceed in a direct way, but by 
$\Upsilon$(5$S$)$\rightarrow$$B^{(*)}$$\overline{B}^{(*)}$ 
and subsequent 
$B^{(*)}$$\overline{B}^{(*)}$$\rightarrow$$\Upsilon$(1$S$,2$S$,3$S$)$\pi^+$$\pi^-$.
On the one hand, the peak position could be shifted upwards 
by the rescattering by +(7$-$20)~MeV \cite{yb_meng_chao}, compatible with the observed
higher peak position of the Y$_b$(10889) compared to the $\Upsilon$(5$S$). 
On the other hand, the amplitude of the rescattering is proportional
to $|$$\vec{p}_1$$|^3$, where $\vec{p}_1$ denotes the 3-momentum
of the $B^{(*)}$ or $\overline{B}^{(*)}$, and would lead to an anhancement
of the observed cross section by a factor 200$-$600 \cite{yb_meng_chao}. 
This way this mechanism could also provide an explanation for the observed anomalous yield (see above). 
Quantitative predictions are however difficult, because 
unknown form factors \cite{yb_simonov} \cite{yb_meng_chao} must be assumed.

\begin{figure}[t!]
\begin{center}
\centerline{\includegraphics[width=0.7\textwidth]{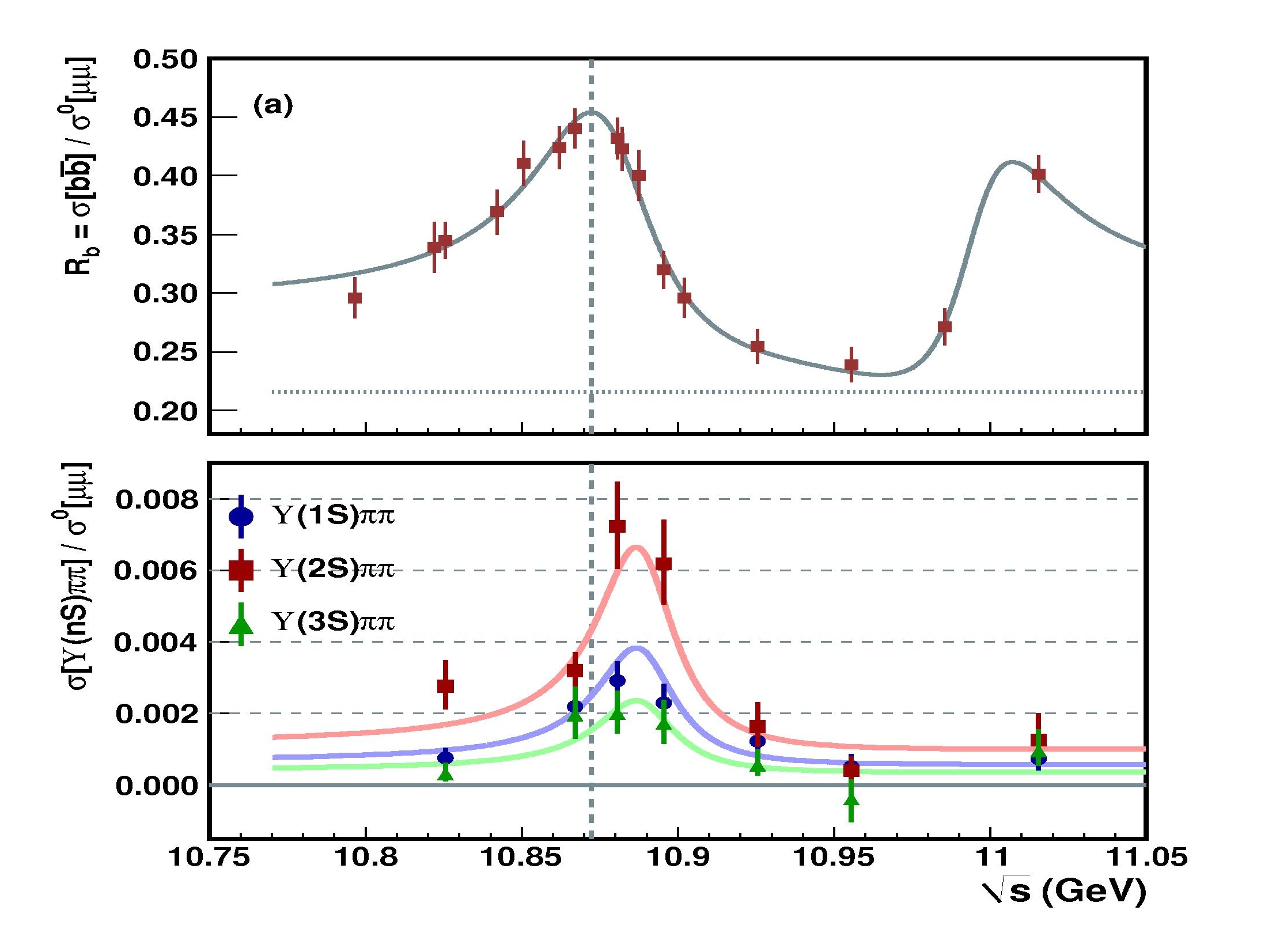}}
\end{center}
\caption{$R_b$ as a function of $\sqrt{s}$ {\it (top)} and the energy-dependent cross sections 
for $e^+$$e^-$$\rightarrow$$\Upsilon(nS)$$\pi^+$$\pi^-$ ($n=1,2,3$) processes {\it (bottom)}.
The results of the fits are shown as smooth curves.
The vertical dashed line indicates the mass of the $\Upsilon$(5$S$), as determined 
from the fit in the upper plot (i.e.\ the measured location of the maximum hadronic cross section).
\label{fyb_scan}}
\end{figure}

\begin{table}[hhh]
\begin{center}
\begin{tabular}{|l|l|l|l|}
\hline
Process & $\Gamma_{}$ & $\Gamma_{e^+e^-}$ & $\Gamma_{\Upsilon(1S)\pi^+\pi^-}$\\
\hline 
\hline 
$\Upsilon$(2$S$)$\rightarrow$$\Upsilon$(1$S$)$\pi^+$$\pi^-$ & 0.032 MeV & 0.612 keV & 0.0060 MeV \\
\hline 
$\Upsilon$(3$S$)$\rightarrow$$\Upsilon$(1$S$)$\pi^+$$\pi^-$ & 0.020 MeV & 0.443 keV & 0.0009 MeV \\
\hline 
$\Upsilon$(4$S$)$\rightarrow$$\Upsilon$(1$S$)$\pi^+$$\pi^-$ & 20.5 MeV & 0.272 keV & 0.0019 MeV \\
\hline 
$\Upsilon$(10860)$\rightarrow$$\Upsilon$(1$S$)$\pi^+$$\pi^-$ & 110 MeV & 0.31 keV & 0.59 MeV \\
\hline 
\end{tabular}
\end{center}
\caption{Total widths, partial width for decay into $e^+$$e^-$ and 
partial width for decay into $\Upsilon$(1$S$)$\pi^+$$\pi^-$ for the 
$\Upsilon$(2$S$), $\Upsilon$(3$S$) and $\Upsilon$(5$S$). The $\Upsilon$(5$S$)
is denoted as $\Upsilon$(10860), as it might be an admixture of several
closeby states. As can be seen, $\Gamma_{\Upsilon(1S)\pi^+\pi^-}$ 
is anomolously large by a factor $>$10$^2$ for the $\Upsilon$(10860).\label{tyb10889}}
\end{table}

\section{A future Project: measurement of the width of the X(3872)}

One of the important steps would be to measure not only the {\it masses}
of newly observed states, but also the {\it widths}. 
As many states have natural widths in the sub-MeV regime, 
future experiments must be able to reach according precision.
The \panda experiment at FAIR (Facility for Antiproton and Ion
research) at GSI Darmstadt, Germany, 
will be using a stored, cooled anti-proton beam. 
The measurement of the width of a state can be performed by a resonance scan technique. 
Both stochastic cooling and $e^-$-cooling techniques will be used,
providing a momentum resolution of the antiproton beam of down to
$\Delta$$p$/$p$$\geq$2$\times$10$^{-5}$.
The anti-protons will collide with protons in e.g.\ a frozen pellet target.
With a maximum beam momentum of $p$$\leq$15~GeV/c, 
in this fixed target setup 
a maximum center-of-mass energy of $\sqrt{s}$$\leq$5.5~GeV can be achieved,
corresponding to a very high mass of an accessible charmonium(-like) state,
which would kinematically not be accessible in $B$ meson decays or in radiative decays
of $\psi$ resonances. 
For momentum reconstruction, a high magnetic solenoid field of $B$=2~T 
will be employed. 
One of the difficulties will be, that signal events (e.g.\ charmonium production, 
with subsequent decays into light mesons) 
and background events (hadronic production of light mesons)
have very similar topologies. Thus, a hardware trigger using simple criteria, 
such as number of charged tracks or number of photons in the calorimeter,
is not possible. Therefore \panda will perform complete online reconstruction
of all events with a high interaction rate of $\leq$2$\times$10$^7$/s.
The planned luminosity of {\cal L}=2$\cdot$10$^{32}$~cm$^{-2}$~s$^{-1}$ 
is high and would translate
into a number of 2$\cdot$10$^9$ $J$/$\psi$ per year, 
if theoretically running on the $J$/$\psi$ resonance only. 

Cross sections in $p$$\overline{p}$ formation (as an example
$\sigma$($p$$\overline{p}$$\rightarrow$X(3872))
can be estimated
from measured branching fractions
(i.e.\ ${\cal B}$(X(3872)$\rightarrow$$p$$\overline{p}$)
using the principle of detailed balance,
which is shown in Eq.~\ref{edetailed_balance}.

\begin{eqnarray}
\sigma[ p\overline{p} \rightarrow X(3872) ] & = &
\sigma_{BW}[ p\overline{p} \rightarrow X(3872) \rightarrow {\rm all}](m_{X(3872)}) \nonumber\\
& =  & \frac{(2J+1) \cdot 4\pi}{m_{X(3872)}^2 - 4 m_p^2} \cdot
\frac{{\cal B}(X(3872)\rightarrow p\overline{p}) \cdot
\overbrace{{\cal B}(X(3872)\rightarrow f)}^{=1} \cdot \Gamma_{X(3872)}^2}
{\underbrace{4(m_{X(3872)} - m_{X(3872)})^2}_{=0} + \Gamma_{X(3872)}^2} \nonumber \\
&\stackrel{(J=1)}{=} & \frac{3\cdot 4\pi}{m_{X(3872)}^2-4 m_p^2}
\cdot {\cal B}(X(3872)\rightarrow p\overline{p}) \ .
\label{edetailed_balance}
\end{eqnarray}

\begin{table*}[htb]
\begin{center}
\begin{tabular}{|l|l|l|l|l|l|}
\hline
$R$ & $J$ & $m$ $[$MeV$]$ &  $\Gamma$ $[$keV$]$ & ${\cal B}$($R$$\rightarrow$$p$$\overline{p}$) & $\sigma$($\overline{p}$$p$$\rightarrow$$R$) \\
\hline
$J$/$\psi$& 1 & 3096.916$\pm$0.011 & 92.9$\pm$2.8 & (2.17$\pm$0.07)$\times$10$^{-3}$ & 5.25$\pm$0.17~$\mu$b\\
$\psi'$ & 1 & 3686.109$^{+012}_{-014}$ & 304$\pm$9 & (2.76$\pm$0.12)$\times$10$^{-4}$ & 402$\pm$18~nb\\
$\eta_c$ & 0 & 2981.0$\pm$1.1 & (29.7$\pm$1.0)$\times$10$^3$ & (1.41$\pm$0.17)$\times$10$^{-3}$ & 1.29$\pm$0.16~$\mu$b\
\\
$\eta_c'$ & 0 & 3638.9$\pm$1.3 & (10$\pm$4)$\times$10$^3$ & (1.85$\pm$1.26)$\times$10$^{-4}$ & 93$\pm$63~nb\\
$\chi_{c0}$ & 0 & 3414.75$\pm$0.31 & (10.4$\pm$0.6)$\times$10$^3$ & (2.23$\pm$0.13)$\times$10$^{-4}$ & 134.1$\pm$7.8\
~nb\\
$h_c$ & 1 & 3525.41$\pm$0.16 & $\leq$1$\times$10$^3$ & (8.95$\pm$5.21)$\times$10$^{-4}$ & 1.47$\pm$0.86~$\mu$b\\
X(3872) & 1 & 3871.68$\pm$0.17 & $\leq$1.2$\times$10$^3$ & $\leq$5.31$\times$10$^{-4}$ & $\leq$68.0~nb\\
\hline
\end{tabular}
\end{center}
\caption{Total spin $J$, mass $m$, width $\Gamma$, branching fraction for the decay into $p$$\overline{p}$
and cross sections for production at {${\sf \overline{P}ANDA}$}, 
as derived by the principle of detailed balance for selected resonances $R$.
\label{tdetailed_balance}}
\end{table*}

Tab.~\ref{tdetailed_balance} summarizes cross sections for production at \panda as derived
by the principle of detailed balance for selected resonances $R$.
For the $J$/$\psi$, the $\psi'$, the $\eta_c'$ and the $\chi_{c0}$ the branching fraction
{\cal B}($R$$\rightarrow$$p$$\overline{p}$) was taken from \cite{pdg}.
For the $\eta_c'$,
{\cal B}($b$$\rightarrow$$K^+$$R$$\rightarrow$$K^+$$p$$\overline{p}$) was taken from \cite{lhcb_ppbark}
and {\cal B}($B^+$$\rightarrow$$K^+$$R$) was taken from \cite{pdg}.
For the $h_c$ and the X(3872)
{\cal B}($b$$\rightarrow$$K^+$$R$$\rightarrow$$K^+$$p$$\overline{p}$) was taken from \cite{lhcb_ppbark}
and the upper limit for {\cal B}($B^+$$\rightarrow$$K^+$$R$) was taken from \cite{pdg}.
Typical cross sections for charmonium formation at \panda are thus in the order
of 10-100~nb. In the following, we assume 
$\sigma$($p$$\overline{p}$$\rightarrow$X(3872))=50~nb. 

Detailed Monte-Carlo simulation studies of a resonance scan 
for $p$$\overline{p}$$\rightarrow$X(3872) at \panda were performed. 
The advantage is, that in $p$$\overline{p}$ collisions the X(3872)
with $J^{PC}$=1$^{++}$ can be formed directly, while in $e^+$$e^-$
only $J^{PC}$=1$^{--}$ is possible. 
The Breit-Wigner cross section for the formation and subsequent decay of a 
$c$$\overline{c}$ resonance $R$ of spin $J$, mass $M_R$ and total 
width $\Gamma_R$ formed in the reaction $\overline{p}$$p$$\rightarrow$$R$ is

\begin{equation}
\sigma_{BW} ( E_{cm} ) =
\frac{(2J+1)}{(2S+1)(2S+1)}
\frac{4 \pi (\hbar c)^2}{ ( E_{cm}^2 - 4 (m_p c^2)^2 )}
\times 
\frac{\Gamma_R^2 BR(\overline{p} p\rightarrow R) \times BR(R\rightarrow f)}
{(E_{cm} - M_R c^2 )^2 + \Gamma_R^2/4}
\end{equation}

where $S$ is the spin of the (anti-)proton.

\begin{equation}
\sigma ( E_{cm} ) =
\int_0^{\infty} 
\sigma_{BW} ( E' ) G (E' - E_{cm} ) dE'
\end{equation}

is a convolution of a Breit-Wigner term for the resonance
and the function $G$ for the beam resolution.
If $G$ is given by a single Gaussian distribution,
then the convolution is a Voigtian distribution.
The area under the resonance peak is given by

\begin{equation}
A = \int_0^{\infty} 
\sigma ( E_{cm} ) dE_{cm} =
\frac{\pi}{2} \sigma_{peak} \Gamma_{R}
\label{earea}
\end{equation}

which importantly is independent of the form of $G$($E$).
$\sigma_{peak}$ is the cross section at $E_{cm}$=$M_Rc^2$ given by

\begin{equation}
\sigma_{peak} =
\frac{(2J+1)}{(2S+1)(2S+1)}
\frac{16 \pi \hbar^2  
BR(\overline{p} p\rightarrow R) \times BR(R\rightarrow f) 
}{
( M_R - 4 m_p^2 ) c^2 \quad .
}
\end{equation}

By measuring $A$ using a fit to the excitation function 
and inserting $\sigma_{peak}$ into Eq.~\ref{earea}, 
the resonance width $\Gamma_R$ can be determined.
For a complete simulation of the resonance scan, 20 simulations for
$p$$\overline{p}$$\rightarrow$X(3872)$\rightarrow$$J$/$\psi$$\pi^+$$\pi^-$
with background were performed for 20 beam momenta in the resonance region.
The beam momenta were chosen equidistant in center-of-mass energy.
For each scan point, the yield of the X(3872) was fitted by a single Gaussian. 
Fig.~\ref{fpandascanfit} shows the fitted yield as a function of $\sqrt{s}$.
The fit was performed using a Voigtian distribution.
Direct background from $p$$\overline{p}$$\rightarrow$$J$/$\psi$$\pi^+$$\pi^-$
was taken into account as a zeroth order polynomial, 
although estimates \cite{chen_ma} indicate that it is small with a cross section of 1.2~nb
(i.e.\ a factor $\simeq$40 smaller than the signal).
The known momentum resolution in the HESR high resolution mode was fixed as the width of the Gaussian in the
convoluted Voigtian.
The width of the X(3872) was reconstructed as $\Gamma_{X(3872}$=86.9$\pm$16.8~keV,
which is consistent with the input width of 100~keV.
This simulation is a proof for the concept, and the ability of \panda to measure 
the width of a resonance in the sub-MeV regime. 
For additional details see \cite{soeren_paper03} \cite{master_martin}.

\begin{figure}
\centering
\includegraphics[width=0.6\textwidth]{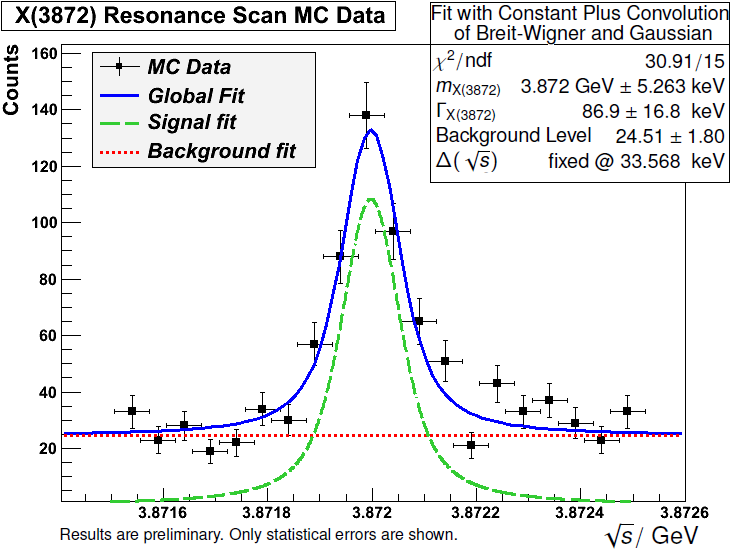}
\caption{Final result for the simulated resonance scan of X(3872) at \panda
with 20 scan points. For details see \cite{master_martin}.}
\label{fpandascanfit}
\end{figure}

\section{Summary}

Recent results from $e^+$$e^-$ collisions (and in particular the $B$ meson factories) 
enable unique precision tests of the $q$$\overline{q}$ potential
in the charmonium and bottomonium region. 
The static potential model fails for many newly observed states (called XYZ states), indicating
non-$q$$\overline{q}$ phenomena such as possibly tetraquark states, charmed meson molecular states or 
hybrid states. Future experiments such as \panda will provide 
precision tests not only of masses, but also widths in the sub-MeV regime. 

\section{Acknowledgments}

The author is grateful for the invitation, the hospitality in Dubna and many interesting and 
inspiring discussions during the school.

\begin{footnotesize}

\bibliographystyle{unsrt}
\bibliography{lange}
 
\end{footnotesize}

\end{document}